\documentclass[showpacs,prb,preprint]{revtex4}
\usepackage{amssymb}
\usepackage{graphicx}

\begin{document}

\title{Magnetoresistance in bilayer graphene via ferromagnet proximity effects}

\author{Y. G. Semenov, J. M. Zavada, and K. W. Kim}
\address{Department of Electrical and Computer Engineering, North
Carolina State University, Raleigh, NC 27695-7911}

\begin{abstract}
A drastic modification of electronic band structure leading to the
magnetoresistance is predicted in bilayer graphene when it is
placed between two ferromagnetic insulators. Due to the exchange
interaction with the proximate ferromagnet, the electronic energy
dispersion in the graphene channel strongly depends on the
magnetization orientation of two ferromagnetic layers,
$\mathbf{M_{1}}$ and $\mathbf{M_{2}} $. While the parallel
configuration $\mathbf{M_{1}}= \mathbf{M_{2}}$ leads to simple
spin splitting of both conduction and valence bands, an energy gap
is induced as soon as the angle $\theta$ between $\mathbf{M_{1}}$
and $ \mathbf{M_{2}}$ becomes non-zero with the maximum achieved
at $\theta=\pi$ (i.e., antiparallel alignment). Consequently,
bilayer graphene may exhibit a sizable magnetoresistance effect in
the current-in-plane configuration. A rough estimate suggests that
the resistance change on the order of tens of percent is possible
at room temperature. This effect is expected to become more
pronounced as the temperatures decreases.
\end{abstract}

\pacs{73.21.-b,85.75.-d,73.43.Qt,73.61.Wp}

\maketitle

With the advent of free-standing atomically thin graphite films
(or graphene),~\cite{Novoselov04} unusual properties related to
the two-dimensional Dirac-like relativistic spectrum of the
honeycomb carbon lattice~\cite{Semenoff84} put graphene in the
forefront of emerging carbon based electronics. Among others, the
half-integer quantum Hall
effect~\cite{Novoselov05,Zhang05,Gusynin05} observed even at room
temperature,~\cite{Novoselov07} high carrier mobility, easy
control of electron and hole concentrations via variation of
applied bias, absence of weak localization and universal minimal
conductivity at the Dirac point with zero density of states have
attracted significant theoretical and experimental attention to
electronic transport in graphene.~\cite{Geim07,Falko07,Neto07}

The spin dependent properties of graphene also offer fascinating
opportunities.  In most studies, the starting point of
consideration is its extremely small spin-orbital coupling
compared to typical semiconductors.~\cite{Min06,Hernando06}
Consequently, graphene exhibits long electron spin-relaxation
time~\cite{Huertas07} and mean free paths~\cite{Oezylmaz07,Cho07}
even at room temperature.~\cite{Tombros07}
%Similarly, it might be a material
%for spin qubits and quantum computation considering the low
%hyperfine interaction of the localized electron spin with the
%$^{13}$C carbon isotopes.~\cite{Loss07}
However, this very advantage (i.e., the weak spin-orbital
interaction) presents a severe challenge in spin manipulation or
selection via electrical control. One possible approach may be to
utilize the electron exchange interaction with proximate
ferromagnetic layers and the resulting effective magnetic
field.~\cite{SKZ07,Brataas07}

It was found very recently that bilayer graphene (BLG) offers a
host of novel phenomena through external modification of the
energy band structure.~%
\cite{Novoselov06,Fal'ko06,Castro06,McDonald07,Cann07}
Particularly, it was shown that a gap opens up between the
conduction and valence energy bands when a potential difference
$u$ is introduced in the two graphene layers. Moreover, the
parabolic band structure near the non-equivalent $K$ and
$K^{\prime }$ points~\cite{Dresselhause02} transforms to a
Mexican-hat-like dispersion with $u \neq 0$.~\cite{Fal'ko06}
Likewise, a non-trivial response may be expected in BLG to the
symmetry-breaking {\em spin-dependent} interactions.

In this study, the properties of BLG sandwiched between two
insulating/dielectric ferromagnets with magnetic moments
$\mathbf{M}_{1}$ and $\mathbf{M}_{2}$ are exploited based on an $8
\times 8$ tight-binding model. Similar to the electrical bias, the
calculation predicts a significant modification of the electronic
band structure when the constituting graphene layers are subject
to un-equal exchange interactions with the proximate magnetic
ions. A particularly interesting feature appears for ferromagnetic
layers with identical magnitude of the magnetic moments
($\mathrm{M}_{1}=\mathrm{M}_{2}$), where $u=0$ and the
\textit{potential asymmetry} is not a factor. As the alignment of
$\mathbf{M}_{1}$ and $\mathbf{M}_{2}$ deviates from each other,
the spin interactions reveal an asymmetry that can alter the BLG
energy bands (including an energy gap) and subsequently the
in-plane conductivity.  The resulting magnetoresistance effect can
play an important role in the carbon based spintronics. In the
case of monolayer graphene, on the other hand, the net effect of
two magnetic layers simply reduces to electron spin splitting in
the effective field induced by the vectorial sum
$\mathbf{M}_{1}+\mathbf{M}_{2}$.

Figure~1 schematically illustrates the specific structure under
consideration. It resembles the ferromagnet-metal hybrid
structures~\cite{Fert88} that reveal a giant magnetoresistance
owing to the spin-dependent conductivity. The bottom ferromagnetic
dielectric layer (FDL) possesses the magnetization $
\mathbf{M}_{1}$ that can be pinned along the direction of the $x$
axis by an antiferromagnetic substrate. The top FDL may be
constructed from the same material but its magnetization vector
$\mathbf{M}_{2}$ can be rotated on the $x$-$y$ plane (by an
external magnetic field) forming an angle $\theta $ with $
\mathbf{M}_{1}$. The influence of FDL magnetization on BLG
electronic structure can be realized in actual structures through
either the direct exchange interaction with magnetic ions
(assuming an overlap between the carbon $\pi $-orbitals and
unfilled shells of the magnetic ions in FDLs) or an indirect
interaction via the ligands of FDLs. Thus, the problem can be
modeled in the mean field approximation with the Hamiltonian
\begin{equation}
H=H_{BL}+\mathcal{P}_{1}\alpha \mathbf{M}_{1}\mathbf{S+}\mathcal{P}%
_{2}\alpha \mathbf{M}_{2}\mathbf{S} \,, \label{f1}
\end{equation}%
where $H_{BL}$ is the spin-independent BLG Hamiltonian.  Two
remaining terms of Eq.~(\ref{f1}) describe the energy of an
electron spin $\mathbf{S}$ in the effective fields (in units of
energy) $\alpha \mathbf{M}_{1}$ and $\alpha \mathbf{M}_{2}$ of the
proximate FDLs. Accordingly, projection operator $\mathcal{P}_{1}$
($\mathcal{P}_{2}$) is 1 for the electron localized at the bottom
(top) carbon monolayer and 0 otherwise. Parameter $\alpha $ is
proportional to the carrier-ion exchange constant as evaluated in
Refs.~\onlinecite{SKZ07} and \onlinecite{Brataas07}.

In the case of low energy electronic excitations, the
tight-binding approximation~\cite{Fal'ko06} can accurately
describe the BLG band spectra as recently demonstrated in a
density functional calculation.~\cite{McDonald07} Hence, we adopt
the tight-binding Hamiltonian near the valley extrema in a basis
that constitutes the components ($
A_{1}\uparrow $, $A_{1}\downarrow $, $B_{2}\uparrow $, $B_{2}\downarrow $, $%
A_{2}\uparrow $, $A_{2}\downarrow $, $B_{1}\uparrow $, $B_{1}\downarrow $)
for the $K$ valley and ($B_{2}\uparrow $, $B_{2}\downarrow $, $A_{1}\uparrow
$, $A_{1}\downarrow $, $B_{1}\uparrow $, $B_{1}\downarrow $, $A_{2}\uparrow $%
, $A_{2}\downarrow $) for the $K^{\prime }$ valley.  Here,
$A_{i}$ and $ B_{i}$ correspond to the electron amplitudes at
inequivalent sites of the bottom ($i=1$) and top ($i=2$) graphene
layers as shown in Fig.~1(b), and $\uparrow $ and $%
\downarrow $ denote spin up and spin down states.

The Hamiltonian $H_{BL}$ includes the lateral coupling for nearest
carbon atoms in the bottom ($A_{1}-B_{1}$) and top ($A_{2}-B_{2}$)
layers with the matrix element $\gamma$ ($ =3$ eV) as well as the
interlayer coupling for $A_{2}-B_{1}$ and $ A_{1}-B_{2}$ dimers
with the matrix elements $\gamma _{1}$ ($=0.4$ eV) and $\gamma
_{3}$ ($=0.3$ eV), respectively; hence, $\gamma _{3}<\gamma
_{1}\ll \gamma $. At zero magnetic field, Eq.~(\ref{f1}) for the
lowest electronic states in the $K$ and $K^{\prime }$ valleys can
be expressed in terms of the in-plane electron momentum
$\mathbf{k}= (k_{x},k_{y})$ defined from the centrum of each
valley  as
\begin{equation}
H_{\varkappa }=\varkappa \left(
\begin{array}{cccccccc}
-\frac{1}{2}u & \frac{\varkappa }{2}\alpha M_{\varkappa } &
v_{3}ke^{i\varphi } & 0 & 0 & 0 & vke^{-i\varphi } & 0 \\
\frac{\varkappa }{2}\alpha M_{\varkappa }^{\ast } & -\frac{1}{2}u & 0 &
v_{3}ke^{i\varphi } & 0 & 0 & 0 & vke^{-i\varphi } \\
v_{3}ke^{-i\varphi } & 0 & \frac{1}{2}u & \frac{\varkappa }{2}\alpha
M_{-\varkappa } & vke^{i\varphi } & 0 & 0 & 0 \\
0 & v_{3}ke^{-i\varphi } & \frac{\varkappa }{2}\alpha M_{-\varkappa }^{\ast }
& \frac{1}{2}u & 0 & vke^{i\varphi } & 0 & 0 \\
0 & 0 & vke^{-i\varphi } & 0 & \frac{1}{2}u & \frac{\varkappa }{2}\alpha
M_{-\varkappa } & \varkappa \gamma _{1} & 0 \\
0 & 0 & 0 & vke^{-i\varphi } & \frac{\varkappa }{2}\alpha M_{-\varkappa
}^{\ast } & \frac{1}{2}u & 0 & \varkappa \gamma _{1} \\
vke^{i\varphi } & 0 & 0 & 0 & \varkappa \gamma _{1} & 0 & -\frac{1}{2}u &
\frac{\varkappa }{2}\alpha M_{\varkappa } \\
0 & vke^{i\varphi } & 0 & 0 & 0 & \varkappa \gamma _{1} & \frac{\varkappa }{2%
}\alpha M_{\varkappa }^{\ast } & -\frac{1}{2}u%
\end{array}%
\right) ,  \label{f2}
\end{equation}%
where the index $\varkappa $ separates the case of $K$ ($\varkappa
=+1$) and $K^{\prime }$ $(\varkappa =-1)$ valleys, $ \varphi =
\tan^{-1} (k_{y}/k_{x})$, $M_{+1}=M_{1}$, $M_{-1}=M_{2}e^{-i
\theta }$, and $v=\sqrt{3}a\gamma /2\hbar $ is the electron
velocity at the Fermi energy in monolayer
graphene.~\cite{Novoselov05}   In addition, the term $
v_{3}=\sqrt{3}a\gamma _{3}/2\hbar$ ($ \ll v$) is responsible for
the trigonal warping, where $a=0.249$ nm is the length of lattice
unit vector. The secular equation for Eq.~(\ref{f2}) is solved
under the conditions of (i) zero bias and (ii) identical top and
bottom FM materials (i.e., $u=0$, $M_{1}=M_{2}\equiv M$). For
simplicity, the transfer matrix elements for off-center sites
$A_{1}$ and $B_{2}$ are also ignored ($ v_{3}\rightarrow 0$). This
approximation does not change qualitatively the results for the
energies $E>1$ meV as discussed in Ref.~\onlinecite{Cann07}.

The specified conditions leads to the energy spectra of each
valley consisting of eight non-degenerate branches $\varepsilon
_{n}(k)$ that are identical for conduction and valence bands and
isotropic with respect to the valley centrum (i.e., independent of
$\varphi $).  Two spin pair solutions correspond to the excited
states with energies $\left\vert \varepsilon _{n}(k)\right\vert
\gtrsim \gamma _{1}$ that are beyond the current interest.  Hence,
only the remaining four low-energy bands $\varepsilon
_{n}=\varepsilon _{n}(k)$ are considered. As it is convenient to
normalize the parameters in units of $\gamma _{1}$, the
dimensionless momentum $p \equiv vk/\gamma _{1}$ and the exchange
field $\mathbf{G} \equiv \alpha \mathbf{M}/\gamma _{1}$ are
introduced hereinafter. The low energy bands $\varepsilon _{n}=\pm
\gamma _{1}E_{\pm }$
can be expressed as%
\begin{equation}
E_{\pm}=\sqrt{p^{2}+\frac{G^{2}}{4}+\frac{1}{2}\left( 1\pm G\cos
\frac{\theta }{2}-W_{\pm }\right)} \,,  \label{f3}
\end{equation}
where
\begin{equation}
W_{\pm }=\sqrt{\left( 1\pm G\cos \frac{\theta }{2}\right)
^{2}(1+4p^{2})+2p^{2}G^{2}(1-\cos \theta )}  \label{f4}
\end{equation}%
for $0\leqslant \theta \leqslant \pi $.  Two solutions with
$\varepsilon _{n}>0$ correspond to the lowest conduction bands,
while their mirror images with respect to the zero energy describe
the highest valence bands.

The most remarkable outcome of the calculation is the presence of
an energy gap $E_{g}$ between the lowest conduction band and the
highest valence band for $\theta \neq 0 $ as shown in Fig.~2. When
the orientation of $ \mathbf{M}_{1}$ and $\mathbf{M}_{2}$ is in
parallel alignment [$\theta =0$, Fig.~2(a)], the net effect of the
exchange interaction simply lifts the two-fold spin degeneracy
resulting in two pairs of spin-split bands that cross each other
at $p=\sqrt{G(1+G/2)/2}$. However, once they are disaligned, the
electronic bands become of mixed spin character (e.g., with both
parallel and antiparallel components to the $x$ direction).
Subsequent anti-crossing opens up the gap that progressively grows
with $\theta$.  At $\theta =\pi $ [i.e.,
$\mathbf{M}_{1}=-\mathbf{M}_{2}$, Fig.~2(d)], the conduction and
valence bands are merged to form two doubly degenerate states with
the maximal $E_{g}=G/\sqrt{1+G^{2}}$ at
$p=G\sqrt{2+G^{2}}/2\sqrt{1+G^{2}}$.  This spin degeneracy is not
surprising because the equivalence of top and bottom graphene
layers makes the $\mathbf{M}_{1}$ and $\mathbf{M}_{2}$
($=-\mathbf{M}_{1}$) directions indistinguishable [i.e., the
dependence of Eqs.~(\ref{f3}) and (\ref{f4}) on sign in $\pm G$
disappears].  Note that in the case of monolayer graphene, the
effect of two FDLs in a similar configuration
$\mathbf{M}_{2}=-\mathbf{M}_{1}$ gives rise to net cancellation of
spin-dependent band modification.

In general, the bandgap induced in BLG can be expressed in term of
the deviation angle $\theta$ as
\begin{equation}
E_{g}=\frac{G\sin \theta /2}{\sqrt{1+G^{2}+2G\cos \theta /2}} \,.
\label{f5}
\end{equation}%
Evident from this equation, the strength of the effective field
$G$ determines the size of $E_{g}$. If its magnitude is comparable
to the thermal energy (strictly speaking $\gamma _{1} G \gtrsim
k_{B}T $ as $G$ is a normalized quantity), the orientation
dependence in Eq.~(\ref{f5}) may manifest itself through the
variation of electron/hole population.  At the same time, the
carrier velocity in BLG is also strongly affected as indicated by
the flattening of the bands with increasing $\theta$ in Fig.~2.
Clearly the conductivity in the BLG channel can be modulated by
the effective fields from the two ferromagnetic barriers, leading
to a sizable magnetoresistance effect in the current-in-plane
configuration.

The effect of both conduction/valence band separation and
deformation on the magnetoresistance $ R(\theta )$ can be taken
into account in the one-electron approximation in terms of
Kubo-Greenwood formula for conductivity~\cite{Kubo,Greenwood}
\begin{equation}
\sigma _{xx}=(2\pi e)^{2}\hbar \sum_{m,n}\sum_{\mathbf{k},\mathbf{k}^{\prime
}}\left\vert \left\langle m,\mathbf{k}\left\vert v_{x}\right\vert n,\mathbf{k%
}^{\prime }\right\rangle \right\vert ^{2}\left( -\frac{\partial f}{\partial
\varepsilon }\right) _{m,\mathbf{k}}\delta (\varepsilon _{m,\mathbf{k}%
}-\varepsilon _{n,\mathbf{k}^{\prime }}),  \label{f6}
\end{equation}%
where $e$ is the electron charge, $m$ and $n$ the subband indices,
$ v_{x}=dx/dt=i[H_{\varkappa },x]/\hbar $ the velocity operator,
and $f$ the Fermi-Dirac function. The case of ballistic
conductivity is examined at finite temperature.~\cite{Nomura07}
The specific property of interest is the relative change $\sigma
_{xx}(\theta )/\sigma _{xx}(0)$ as a function of $\theta$.
Utilizing $G \ll 1$ in most cases, our analysis reveals that it
depends most sensitively on a single parameter $\gamma_1 G/k_{B}T$
as expected. Figure~3 provides the calculated $\sigma _{xx}(\theta
)/\sigma _{xx}(0)$ as well as the corresponding change in the
electron/hole concentration $n (\theta) / n (0)$ for two different
values of $\gamma_1 G/k_{B}T$. The comparatively weaker decrease
of $n$ (in reference to $\sigma_{xx}$) as $\theta $ rotates
signifies a major contribution of the mobility variation due to
the flattened band structure.

Assuming a sensitive response of $\mathbf{M}_{2}$ to the external
magnetic field, one can define the magnetoresistance as a
difference of the BLG resistance $R(0)\sim 1/\sigma _{xx}(0)$ at
$\mathbf{M}_{1}=\mathbf{M}_{2}$ and $R(\pi )\sim 1/\sigma
_{xx}(\pi )$ at $\mathbf{M}_{1}=-\mathbf{M}_{2}$.  Figure~4
presents the normalized result $ \xi =[R(\pi )-R(0)]/R(0)$ as a
function of $\gamma_1 G/k_{B}T$.  Since this quantity depends on
the position of the electro-chemical potential $\mu $ as well,
four different values of $\left\vert \mu \right\vert /k_{B}T$ are
considered.  All the cases show essentially the same
characteristics with the largest effect exhibited at $\mu =0$
(i.e., $\mu$ at the mid gap).  $ \xi$ also saturates at smaller
effective fields as $\mu$ gets closer to the conduction/valence
band edge.  An important point to note is that the
magnetoresistance effect of {\em tens of percent} is possible once
the exchange field strength $\gamma_1 G$  ($\equiv \alpha M$) is
comparable to the thermal energy. By comparison, the alternative
approach based on spin-valve devices in monolayer graphene reveals
only a feeble magnetoresistance due to the weak dependence of the
graphene conductivity on the electronic details of the
ferromagnetic leads.~\cite{Brey07}

As a practical matter, the strength of the exchange field $G$ is
of major importance. Considering a large variety of potential
FDL/graphene interfaces and the current lack of relevant
information, however, \textit{a priori} evaluation of $G$ may be
possible only in a very rough manner.  Following the approach
presented in Ref.~\onlinecite{Brataas07}, we estimate the exchange
energy for a graphene electron interacting with the nearest
stratum of magnetic ions  as $\alpha M =(n_{2FM}/n_{2C})S_{M}J$,
where $n_{2C} $ and $n_{2FM}$ are the areal concentrations of
graphene carbon atoms and magnetic ions in the FDL, respectively,
$S_{M}$ is the mean value of the magnetic ion spins, and $J$ is
the exchange constant. The latter was found to be $J=15$ meV as
deduced from an experiment in an EuO/Al structure with Curie
temperature $T_{c}=69$ K.~\cite{Roesler94} This constant $J$ may
actually be larger in the case of high temperature ferromagnets
with stronger spin-spin inter-ions interaction. With the provision
that $J\sim T_{c}$, $n_{2FM}/n_{2C}=0.1 - 0.2$, $S_{M}=1 - 2.5$
and $T_{c}=500 - 600$ K, we find $ \gamma_1 G $ $(\equiv \alpha M)
\simeq 15 - 65 $~meV.  Although the accuracy of this estimate is
limited, it nonetheless shows the possibility of prominent
magnetoresistance effect even at room temperature. Lower
temperatures will make its manifestation far more apparent for
easier detection.

It is noted that the idealized model of Eq.~(\ref{f1}) may need to
be expanded for detailed analysis of a specific structure.
Imperfect interfaces between the FDL and BLG can be a source of a
random potential that affects the electronic band structure and
transport properties of graphene.~\cite{Nilsson07} In addition,
any difference between the top and bottom FDLs results in a
non-zero bandgap even at parallel orientation of their magnetic
moments. At the same time, the non-trivial manifestation of
electron-electron interaction~\cite{Nilsson06,Castro07} may
interplay with the studied effects at sufficiently low
temperatures. These factors (and many others)~\cite{Neto07} may
need to be taken into consideration as the situation requires.

%In summary, we examine the possibility to modulate the BLG
%electronic band structure via the exchange interaction with
%proximate FDLs.  The effect is realized when two graphene layers
%are subject to un-equal exchange interactions with the proximate
%magnetic ion.  Hence, as the magnetization orientation of one FDL
%rotates against the other, an energy gap gradually opens up
%between the conduction and valence bands that also become nearly
%flat near the extrema.  Both of these features clearly affect the
%conductivity leading to a sizable magnetoresistance in BLG.

This work was supported in part by the US Army Research Office and
the FCRP Center on Functional Engineered Nano Architectonics
(FENA).

\newpage

\begin{center}
\begin{figure}[tbp]
\includegraphics[scale=.42,angle=0]{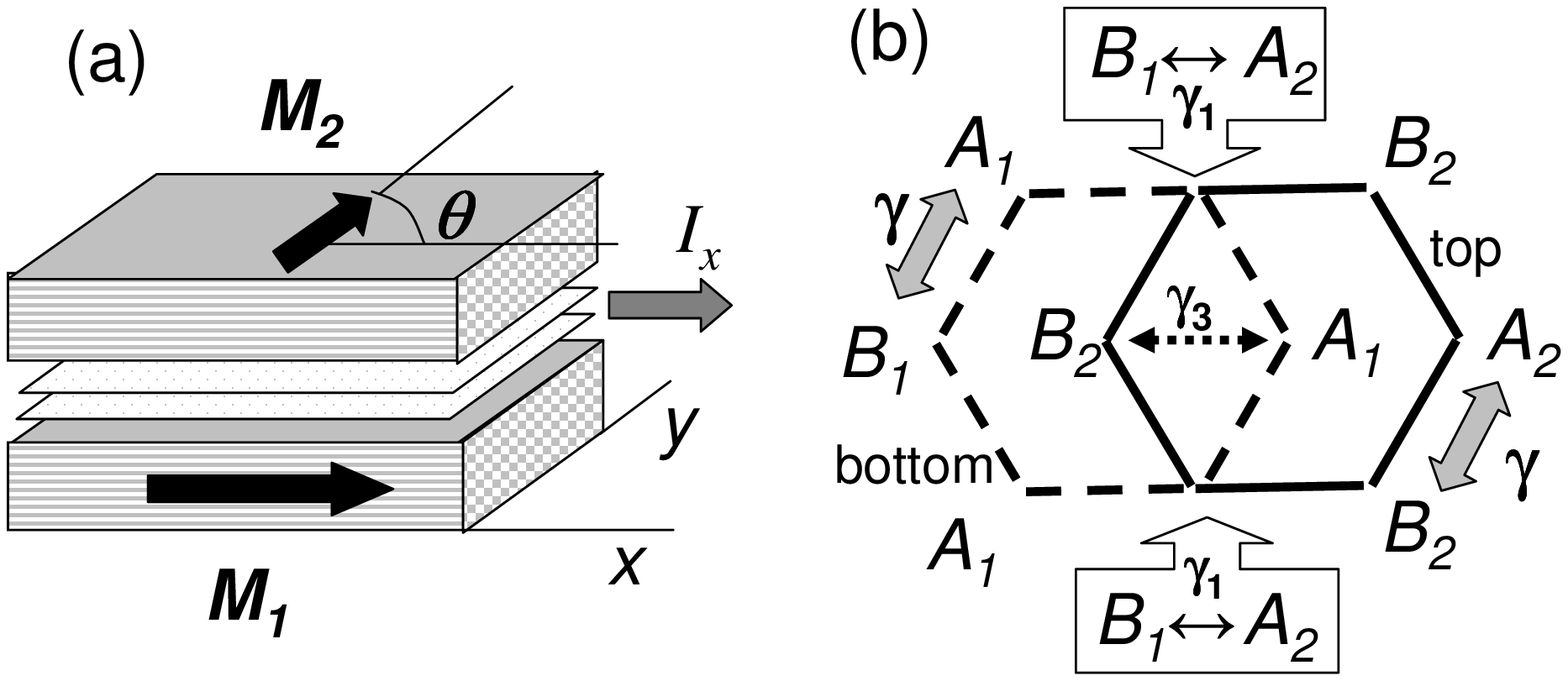}
\caption{(a) Schematic illustration of bilayer graphene (two
closely set planes) sandwiched between ferromagnetic dielectric
layers of magnetization $\mathbf{M}_{1}$ and $\mathbf{M}_{2}$
(separated by the angle $ \protect\theta $). The reference frame
is chosen so that $\mathbf{M}_{1}$ is along the $x$ axis; the grey
arrow shows the probing current $I_{x}$ through the graphene
channel. (b) Fragment of bilayer graphene  as two hexagons (view
from above) with carbon atoms located at the vertices. Lattice
sites $A_{1}$ and $B_{1}$ (dashed hexagon) refer to the bottom
layer, while $A_{2}$ and $ B_{2}$ (solid hexagon) the top layer.
$\protect\gamma$, $\protect\gamma_{1}$, and $\protect\gamma_{3}$
represent the matrix elements for electron transfer between the
nearest in-plane, vertical, and slanted inter-layer carbon atoms,
respectively.}
\end{figure}
\end{center}

\newpage

\begin{center}
\begin{figure}[tbp]
\includegraphics[scale=0.85,angle=0]{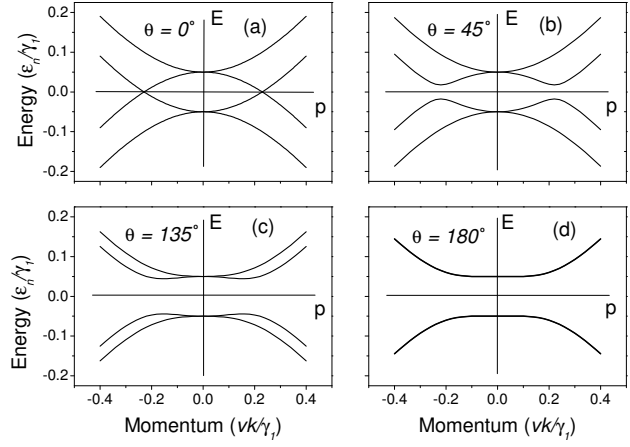}
\caption{Electron energy spectra of bilayer graphene near the $K$
point at different angles $ \protect\theta $ between magnetization
vectors $\mathbf{M} _{1}$ and $\mathbf{M} _{2}$. The reference for
energy $E=0$ corresponds to the center of the bandgap [(b)-(d)] or
the point of contact between the conduction and valence bands in
the case $\protect\theta =0$ [(a)].  The bands are doubly
degenerate at $\protect\theta =\protect\pi $ [(d)].}
\end{figure}
\end{center}

\newpage

\begin{center}
\begin{figure}[tbp]
\includegraphics[scale=0.7,angle=0]{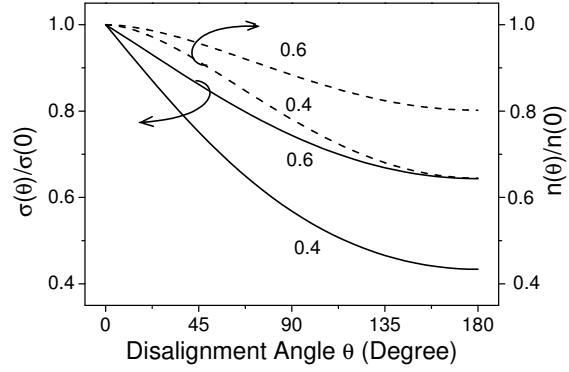}
\caption{Conductivity $\protect\sigma $ (solid lines) and carrier
concentration $n $ (dashed lines) vs.\ $\protect\theta $ in
bilayer graphene at two different ratios of the thermal energy
$k_{B}T$ to the exchange field $\protect\gamma _{1}G$ (i.e.,
$k_{B}T/\protect\gamma _{1}G=$ 0.6 or 0.4). The electro-chemical
potential is fixed at zero (i.e., at the middle of the energy
gap).}
\end{figure}
\end{center}

\newpage

\begin{center}
\begin{figure}[tbp]
\includegraphics[scale=0.7,angle=0]{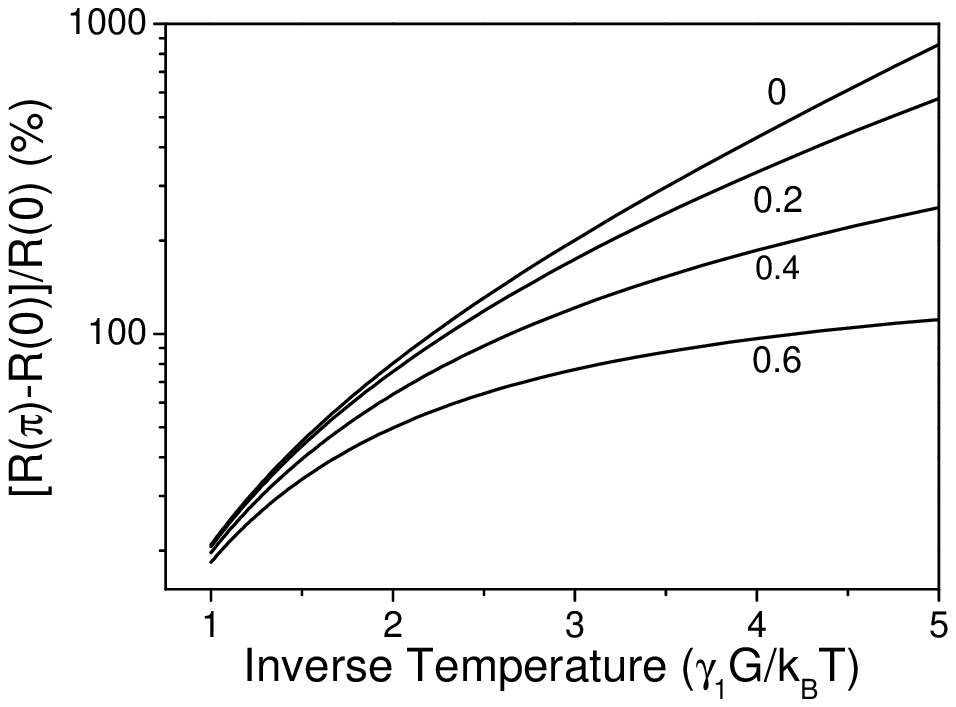}
\caption{Magnetoresistance $\xi$ in the anti-parallel alignment
($\mathbf{M}_{1} = - \mathbf{M} _{2}$) as a function of
$\protect\gamma _{1}G/k_{B}T$ calculated at different locations of
the electro-chemical potential $\protect\mu $. The normalized
value $\protect\mu /k_{B}T$ is provided for each curve.}
\end{figure}
\end{center}


\clearpage

\begin{thebibliography}{99}
\bibitem{Novoselov04} K. S. Novoselov {\it et al.}, Science
\textbf{306}, 666 (2004).

\bibitem{Semenoff84} G. W. Semenoff, Phys. Rev. Lett. \textbf{53}, 2449
(1984).

\bibitem{Novoselov05} K. S. Novoselov {\it et al.}, Nature (London) \textbf{438}, 197 (2005).

\bibitem{Zhang05} Y. Zhang {\it et al.}, Nature (London) \textbf{438}, 201 (2005).

\bibitem{Gusynin05} V. P. Gusynon and S. G. Sharapov, Phys. Rev. Lett.
\textbf{95}, 146801 (2005).

\bibitem{Novoselov07} K. S. Novoselov {\it et al.}, Science \textbf{315}, 1379
(2007).

\bibitem{Geim07} A. K. Geim and K. S. Novoselov, Nat. Mater. \textbf{6}, 183
(2007).

\bibitem{Falko07} V. I. Fal'ko, and A. K. Geim, Eur. Phys. J. \textbf{148},
1 (2007).

\bibitem{Neto07} A. H. Castro Neto {\it et al.}, arXiv:cond-mat/0709.1163 (unpublished).

\bibitem{Min06} H. Min {\it et al.}, Phys. Rev. B \textbf{74}, 165310 (2006).

\bibitem{Hernando06} D. Huertas-Hernando, F. Guinea and A. Brataas, Phys.
Rev. B \textbf{74}, 155426 (2006).

\bibitem{Huertas07} D. Huertas-Hernando, F. Guinea, and A. Brataas, Eur.
Phys. J. \textbf{148}, 177 (2007).

\bibitem{Oezylmaz07} B. Oezylmaz and P. Kim, in \emph{Final Program of the
2007 Electronic Materials Conference} (South Bend, Indiana, 2007), Vol. 1,
p. 85.

\bibitem{Cho07} S. Cho, Y.-Fu Chen, and M. S. Fuhrer, Appl. Phys. Lett.
\textbf{91}, 123105 (2007).

\bibitem{Tombros07} N. Tombros {\it et al.}, Nature (London) \textbf{448}, 571 (2007).

%\bibitem{Loss07} B. Trauzettel, D. V. Bulaev, D. Loss, and G. Burkard,
%Nat. Phys. \textbf{3}, 192 (2007).

\bibitem{SKZ07} Y. G. Semenov, K. W. Kim, and J. Zavada, Appl. Phys. Lett.
\textbf{91}, 153105 (2007).

\bibitem{Brataas07} H. Haugen, D. Huertas-Hernando, and A. Brataas,
arXiv:cond-mat/0707.3976 (unpublished).

\bibitem{Novoselov06} K. S. Novoselov {\it et al.}, Nat. Phys. \textbf{2}, 177 (2006).

\bibitem{Fal'ko06} E. McCann and V. I. Fal'ko, Phys. Rev. Lett. \textbf{96},
086805 (2006).

%\bibitem{Cann06} E. McCann, Phys. Rev. B \textbf{74}, 161403 (2006).

\bibitem{Castro06} E. V. Castro {\it et al.}, Phys. Rev. Lett. \textbf{99}, 216802
(2007).

\bibitem{McDonald07} H. Min {\it et al.}, Phys. Rev. B \textbf{75}, 155115 (2007).

\bibitem{Cann07} E. McCann, D. S. L. Abergel, and V. I. Fal'ko, Eur. Phys. J.
\textbf{148}, 91 (2007).

\bibitem{Dresselhause02} M. S. Dresselhause and D. Dresselhause, Adv. Phys.
\textbf{51}, 1 (2002).

\bibitem{Fert88} See, for example, M. N. Baibich {\it et al.},
Phys. Rev. Lett. \textbf{61}, 2472 (1988).

%\bibitem{Grunberg89} G. Binasch, P. Gr\"{u}nberg, F. Saurenbach, and W.
%Zinn, Phys. Rev. B \textbf{39}, R4828 (1989).

%\bibitem{Parkin90} S. S. P. Parkin, N. More, and K. P. Roche, Phys. Rev.
%Lett. \textbf{64}, 2304 (1990).

%\bibitem{spinel} See, for example, \emph{Spin Electronics\/}, ed.
%D.~Awschalom (Kluwer, Dordrecht, 2004).

%\bibitem{Wolf} S. A. Wolf, A. Y. Chtchelkanova, and D. M. Treger, IBM J.
%Res. Dev. \textbf{50}, 101 (2006).

\bibitem{Kubo} R. Kubo, J. Phys. Soc. Japan \textbf{12}, 570 (1957).

\bibitem{Greenwood} D. A. Greenwood, Proc. Phys. Soc. \textbf{71}, 585
(1958).

\bibitem{Nomura07} The conductance of graphene subjected to a random
potential was evaluated in terms of the Kubo-Greenwood formula by
K. Nomura, M. Koshino, and S. Ryu, Phys. Rev. Lett. \textbf{99},
146806 (2007).

\bibitem{Brey07} L. Brey and H. A. Fertig, Phys. Rev. B \textbf{76}, 205435
(2007).

\bibitem{Roesler94} G. M. Roesler {\it et al.}, Proc. SPIE
\textbf{2157}, 285 (1994).

\bibitem{Nilsson07} J. Nilsson {\it et al.}, arXiv:cond-mat/0712.3259 (unpublished).

\bibitem{Castro07} E. V. Castro {\it et al.}, arXiv:cond-mat/0711.0758 (unpublished).

\bibitem{Nilsson06} J. Nilsson {\it et al.}, Phys. Rev. B \textbf{73}, 214418 (2006).

\end{thebibliography}
\end{document}